\documentclass[twocolumn]{aastex63}

\received{}
\revised{}
\accepted{}
\submitjournal{ApJL}

\shorttitle{An [$\alpha$/Fe]-enhanced thick disk in a Milky Way Analogue}
\shortauthors{Scott et al.}
\graphicspath{{./}{figures/}}

\begin{document}

\title{Identification of an [$\alpha$/Fe]--enhanced thick disk component in an edge-on Milky Way Analogue}

\correspondingauthor{Nicholas Scott}
\email{nicholas.scott@sydney.edu.au}

\author[0000-0002-0786-7307]{Nicholas Scott}
\affiliation{Sydney Institute for Astronomy, School of Physics, University of Sydney, NSW 2006, Australia}
\affiliation{ARC Centre of Excellence for All Sky Astrophysics in 3 Dimensions (ASTRO 3D)}

\author[0000-0003-2552-0021]{Jesse van de Sande}
\affiliation{Sydney Institute for Astronomy, School of Physics, University of Sydney, NSW 2006, Australia}
\affiliation{ARC Centre of Excellence for All Sky Astrophysics in 3 Dimensions (ASTRO 3D)}

\author[0000-0002-0920-809X]{Sanjib Sharma}
\affiliation{Sydney Institute for Astronomy, School of Physics, University of Sydney, NSW 2006, Australia}
\affiliation{ARC Centre of Excellence for All Sky Astrophysics in 3 Dimensions (ASTRO 3D)}

\author[0000-0001-7516-4016]{Joss Bland-Hawthorn}
\affiliation{Sydney Institute for Astronomy, School of Physics, University of Sydney, NSW 2006, Australia}
\affiliation{ARC Centre of Excellence for All Sky Astrophysics in 3 Dimensions (ASTRO 3D)}

\author{Ken Freeman}
\affiliation{Research School of Astronomy \& Astrophysics, Australian National University, ACT 2611, Australia}

\author[0000-0003-3333-0033]{Ortwin Gerhard}
\affiliation{Max-Planck-Institut f\"ur Extraterrestrische Physik, Giessenbachstrasse, D-85741 Garching, Germany}

\author{Michael R. Hayden}
\affiliation{Sydney Institute for Astronomy, School of Physics, University of Sydney, NSW 2006, Australia}
\affiliation{ARC Centre of Excellence for All Sky Astrophysics in 3 Dimensions (ASTRO 3D)}

\author{Richard McDermid}
\affiliation{Research Centre for Astronomy, Astrophysics, and Astrophotonics, Department of Physics and Astronomy, Macquarie University, NSW 2109, Australia}
\affiliation{ARC Centre of Excellence for All Sky Astrophysics in 3 Dimensions (ASTRO 3D)}



\begin{abstract}
The Milky Way disk consists of two prominent components --- a thick, alpha-rich, low-metallicity component and a thin, metal-rich, low-alpha component. External galaxies have been shown to contain thin and thick disk components, but whether distinct components in the [$\alpha$/Fe]--[Z/H] plane exist in other Milky Way-like galaxies is not yet known. We present VLT-MUSE observations of UGC 10738, a nearby, edge-on Milky Way-like galaxy. We demonstrate through stellar population synthesis model fitting that UGC 10738 contains alpha-rich and alpha-poor stellar populations with similar spatial distributions to the same components in the Milky Way. We discuss how the finding that external galaxies also contain chemically distinct disk components may act as a significant constraint on the formation of the Milky Way's own thin and thick disk.
\end{abstract}



\section{Introduction} \label{sec:intro}
The Milky Way is the galaxy for which we have the most detailed observational data, with extremely precise astrometric and kinematic data from the Gaia satellite \citep{Gaia:2016,Gaia:2018}, as well as accurate chemical abundances for large numbers of stars from the APOGEE \citep{Majewski:2017}, GALAH \citep{Buder:2020} and LAMOST/LEGUE \citep{Deng:2012} surveys. Striking features of the Milky Way are the chemically distinct thin and thick disk \citep{Gilmore:1983,Haywood:2013,Hayden:2015}. Thick disk stars are enhanced in [$\alpha$/Fe] with respect to thin disk stars of a similar metallicity, with the overall metallicity lower than that of the thin disk. The two disks have distinct spatial distributions, with the thick disk having a larger scale height of $\sim 900$ pc, compared to the thin disk's $\sim 300$ pc \citep{Juric:2008,Bland-Hawthorn:2016}. 

\begin{figure*}
    \centering
    \includegraphics[width=7in,clip,trim=90 0 90 40]{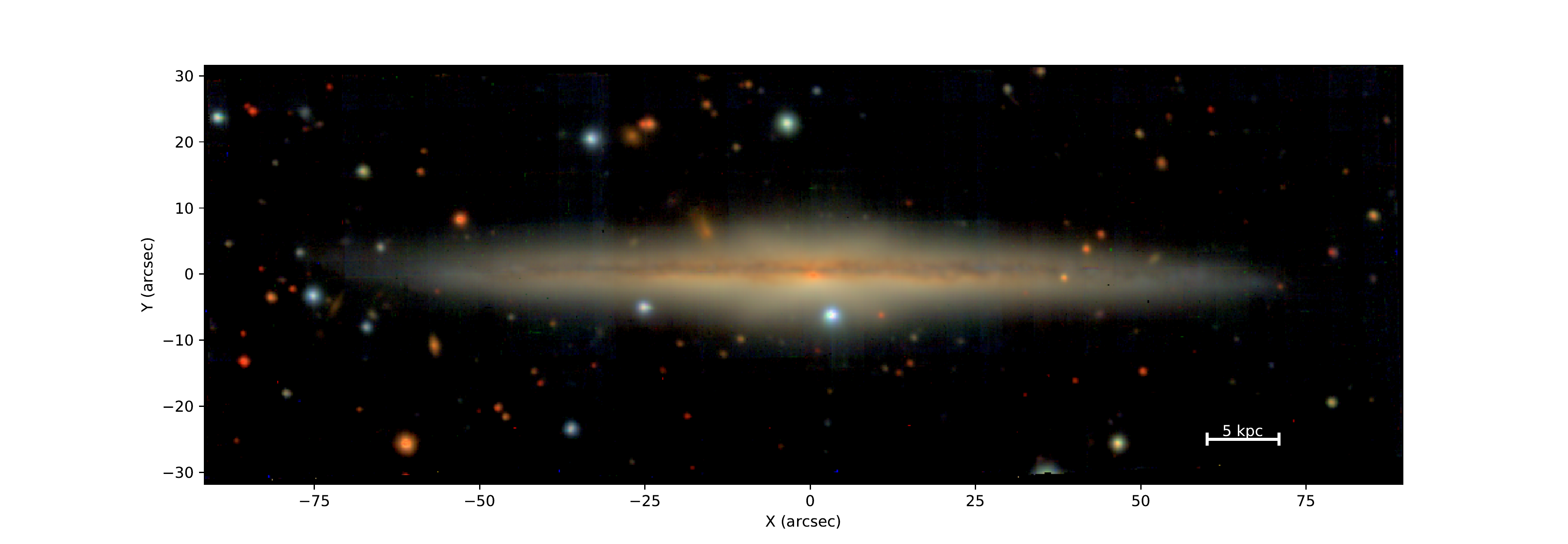}
    \caption{Pseudo-{\it gri} colour image of UGC 10738 created from the MUSE datacube. The X/peanut-shaped bulge and prominent dust lane are clearly visible. The white bar indicates a scale of 5 kpc at the distance of UGC 10738, $D = 99$ Mpc.}
    \label{fig:gri_image}
\end{figure*}

While the Milky Way is the best object to study to understand `Galaxy' evolution, it is challenging to make inferences about galaxy evolution in general. From a sample of one, it is difficult to differentiate between general features of galaxy evolution and aspects specific to the Milky Way's own unique formation history. Structural thick disks have been identified in external galaxies from deep imaging surveys \citep{Yoachim:2006,Comeron:2011,Comeron:2018,Martinez-Lombilla:2019} by applying photometric structural decomposition techniques to images of edge-on galaxies, and are thought to be ubiquitous in disk galaxies \citep{Yoachim:2006,Comeron:2011}. The evidence for chemically distinct thick disks in external Milky Way-like galaxies is inconclusive \citep{Yoachim:2008}, with convincing signatures identified primarily in early-type disk galaxies \citep{Comeron:2016,Pinna:2019,Pinna:2019b,Poci:2019,Poci:2021}, with the exception of a low-mass late-type galaxy in \citet{Comeron:2015}.

This work explores whether chemically distinct thick and thin disks are present in nearby disk galaxies that are structurally similar to our own Galaxy -- Milky Way analogues (MWAs). Recent improvements in integral field spectroscopy enable integrated light measurements that can sample physical scales allowing direct comparison to Milky Way observations. We present Very Large Telescope (VLT) -- Multi Unit Spectroscopic Explorer (MUSE) observations of UGC 10738, a nearby MWA. We derive spatially-resolved star formation and chemical enrichment histories and compare these to similar measurements for the Milky Way. We conclude by discussing the implications of these results for the formation of Milky Way-like galaxies, and for the Milky Way itself.

\section{Data and Observations} \label{sec:sample}

\subsection{UGC 10738}

We selected nearby ($D < 100$\ Mpc) galaxies from the Third Reference Catalogue \citep[RC3,][]{deVaucouleurs:1991} with a morphological type similar to the Milky Way \citep[Sbc or SBbc, Hubble T-type $\sim 4$][]{deVaucouleurs:1978} and a stellar mass $10.4 < M_\star < 11.2 (M_\odot)$. From this sample we selected 9 galaxies in three inclination bins ($45^\circ < i < 60^\circ$, $60^\circ < i < 75^\circ$ and $i > 75^\circ$) and three colour bins based on the ($B-I$) colour, or, when unavailable, multi-band images. UGC 10738 falls in the most inclined and reddest bin.

UGC 10738 has a stellar mass of $8 \times 10^{10}$ M$_\odot$ and a $(g-r)$ colour of 0.95, highly comparable to the Milky Way \citep[][and references therein]{Bland-Hawthorn:2016}. We note that UGC 10738 is larger than the Milky Way by $\sim 50$ per cent in effective radius, though a precise comparison is unfeasible given the very different approaches to measuring size in extragalactic objects and the Milky Way.

Magnitudes and effective radii were derived by fitting a S\'ersic surface brightness profile on Pan-STARRS $gri$ images \citep{Waters:2020} using the code \textsc{ProFound} \citep{2018MNRAS.476.3137R} and \textsc{ProFit} \citep{2017MNRAS.466.1513R}. M$_\star$ was estimated using the $g-i$ colour and $i$ magnitude following \citet{2011MNRAS.418.1587T} and \citet{2015MNRAS.447.2857B} and assuming a \citet{2003ApJ...586L.133C} initial mass function. A pseudo-{\it gri} image, derived from the reduced MUSE datacube is shown in Figure \ref{fig:gri_image}.  

\begin{figure*}
    \centering
    \includegraphics[width=3.5in,clip,trim=70 10 80 50]{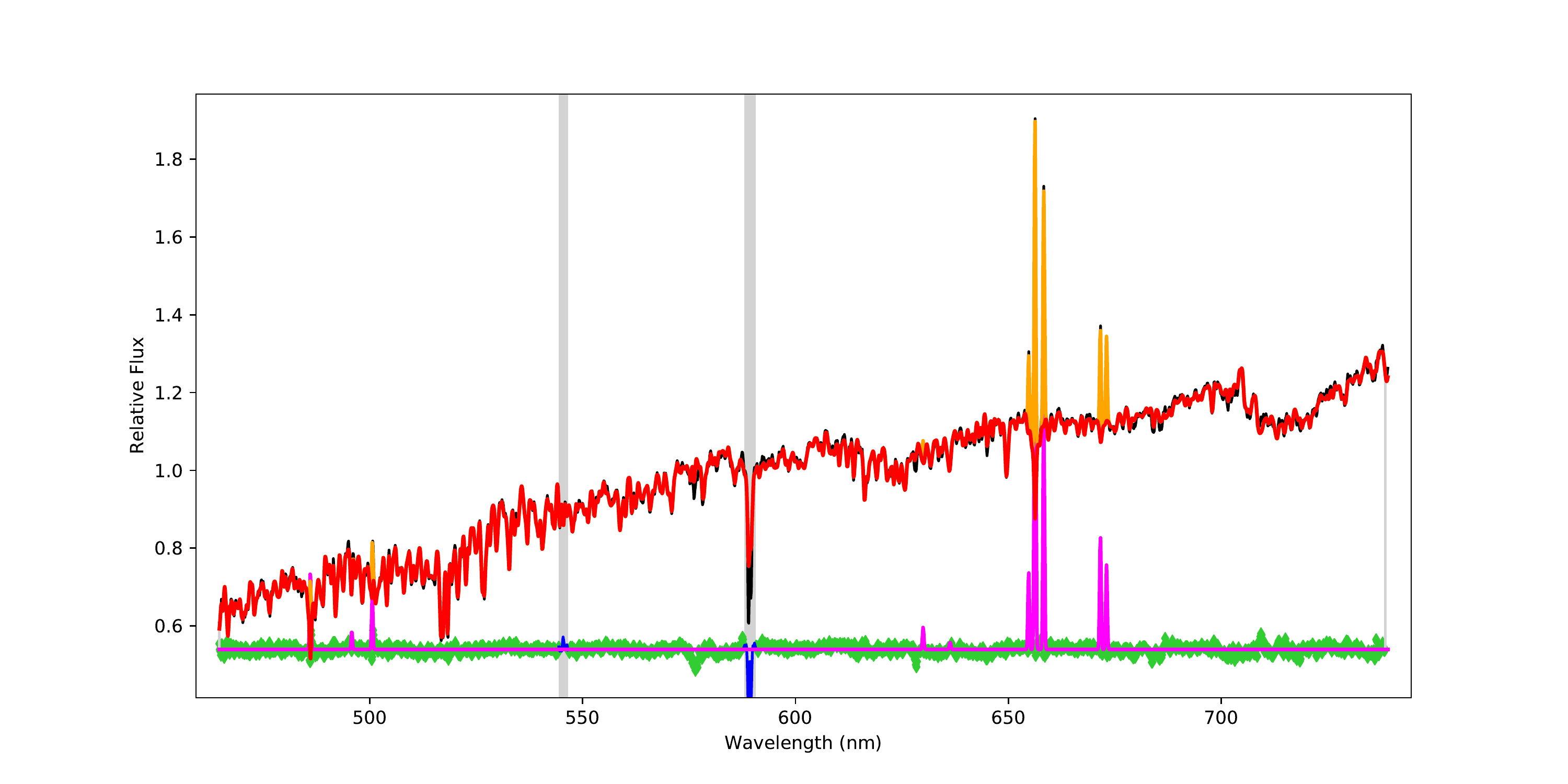}
    \includegraphics[width=3.5in,clip,trim=60 10 80 30]{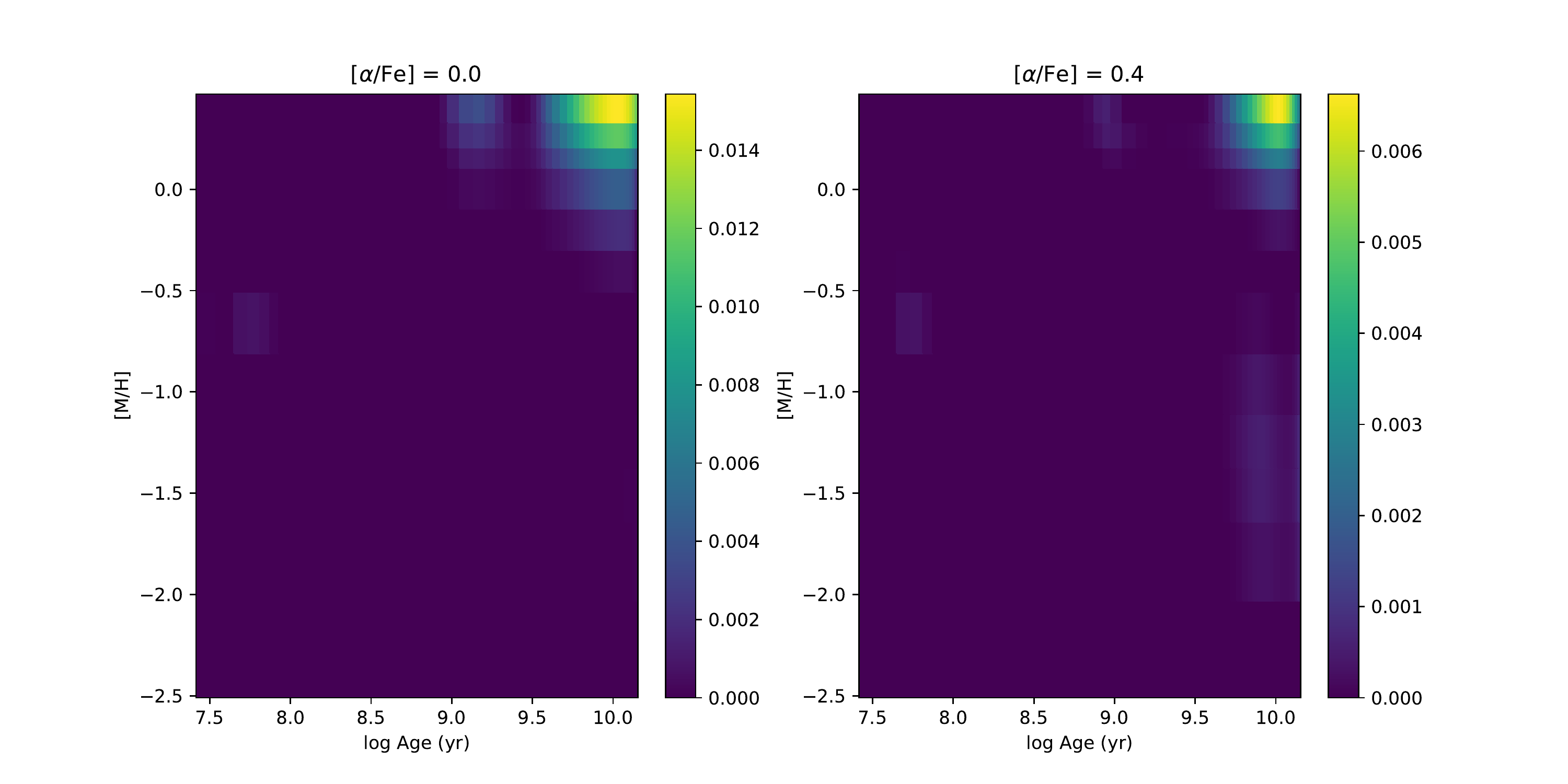}
    \caption{Left panel: example spectral fit with the observed spectrum in black, the stellar component of the fit in red and the gas component in pink and orange. The residuals are in green, with masked regions in blue. Right panels: mass fraction of stars with a given age, [Z/H] and [$\alpha$/Fe] that best reproduce the observed spectrum.}
    \label{fig:pops_example}
\end{figure*}

\subsection{Observations and data reduction}

UGC 10738 was observed with MUSE on the VLT as part of observing program 0101.B-0706(A) (PI: van de Sande) on 16-04-2018 and 17-04-2018 in service mode. MUSE provides a 1\arcmin\ field-of-view with a spatial sampling of 0\farcs2 and spectral coverage from 480 to 930 nm at a resolving power R$\sim 2700$. At our adopted distance, $D = 99$\ Mpc \citep{Tully:2013}, the spatial sampling is 91 pc/spaxel.  

Observations consisted of six 1240 second exposures in three locations, one centred on the centre of the galaxy with the others offset by $\sim$1\arcmin\ in either direction along the major axis of the galaxy, combined with two 90$^\circ$ offset position angles. All observations were reduced using \textsc{pymusepipe}\footnote{https://github.com/emsellem/pymusepipe} version 2.9.9 MUSE Data Reduction Software \citep{2016ascl.soft10004W, 2020A&A...641A..28W} version 2.8.1, including the standard steps of flat fielding, wavelength calibration, flux calibration and telluric correction. Sky subtraction was performed by identifying regions free from galaxy light in each exposure using \textsc{ProFound} \citep{2018MNRAS.476.3137R}, and subtracting the extracted sky spectra from the rest of the data cube using the package \textsc{ZAP} \citep[the Zurich Atmosphere Purge, ][]{2016MNRAS.458.3210S}. The 6 pointings were combined into a single, sky subtracted, flux calibrated mosaic data cube, which was used for the following analysis.

\section{Stellar Population Analysis} \label{sec:stellar_pops}

Our goal in this work is to compare the spatial variation of the [Z/H] and [$\alpha$/Fe] distributions in UGC 10738 to those observed in the Milky Way. We combine spectra into 36 binned regions, chosen to match those used in \citet{Hayden:2015}. Additionally, this greatly increases the signal-to-noise ratio of the binned spectra over the individual spaxels. We define an inner region with $0 < R < 3$\ kpc, and 8 further radial regions each 2 kpc in width, evenly spaced from 3 to 19 kpc along the galaxy major axis. We define four bins in height: $|z| < 0.5$\ kpc, $0.5 < |z| < 1$\ kpc, $1 < |z| < 2$\ kpc and $2 < |z| < 4$\ kpc. These regions extend further than those used in \citet{Hayden:2015}, and as these are projected quantities, the effective intrinsic radii they probe will be relatively larger. Before binning, each spectrum is interpolated onto a common velocity scale based on the velocity map of van de Sande (in prep.) derived with the \textsc{GIST} pipeline \citep[Galaxy IFU Spectroscopy Tool,][]{2019A&A...628A.117B} using the penalized Pixel Fitting (pPXF) \textsc{Python} package \citep{Cappellari:2004} and employing the Voronoi-binning method of \citet{2003MNRAS.342..345C}. Spaxels are co-added from all quadrants of the galaxy to form the final binned spectra for the 18 spatial regions. Spectra covering the region of \citet{Hayden:2015} range in signal-to-noise ratio from $\sim 400$ to $\sim 80$ per spectral pixel, sufficient to recover the distribution of stellar age, [Z/H] and [$\alpha$/Fe] for each region \citep{Liu:2020}.

To measure the star formation and chemical enrichment history we fit the spectra from each region with the $\alpha$-variable SSP library of \citet{Vazdekis:2015}, which is derived from the empirical MILES stellar spectral library \citep{Falcon-Barroso:2011}. The fit is performed over the overlapping wavelength range, 465 to 741 nm. We select SSP templates that form a regular grid with twelve bins in [Z/H] (-2.27, -1.79, -1.49, -1.26, -0.96, -0.66, -0.35, -0.25,  0.06, 0.15,  0.26,  0.4) and fifty three logarithmically spaced bins in age (0.03 to 14 Gyrs). Only two bins in [$\alpha$/Fe] (0.0, 0.4) are used because empirically-based SSP spectral libraries with finer sampling in this dimension were not available at time of publication.

\begin{figure}
    \centering
    \includegraphics[width=3.5in,clip,trim = 10 10 20 10]{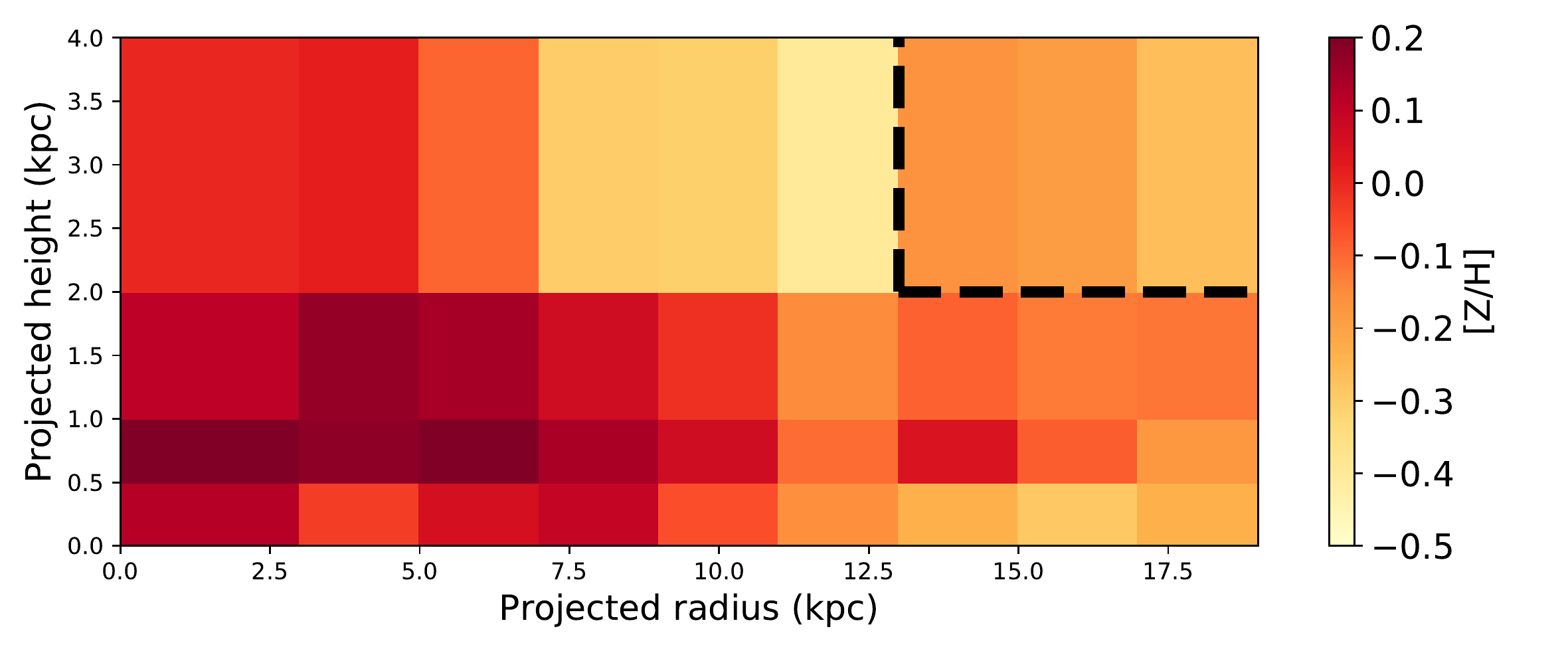}
    \includegraphics[width=3.5in,clip,trim = 10 10 20 10]{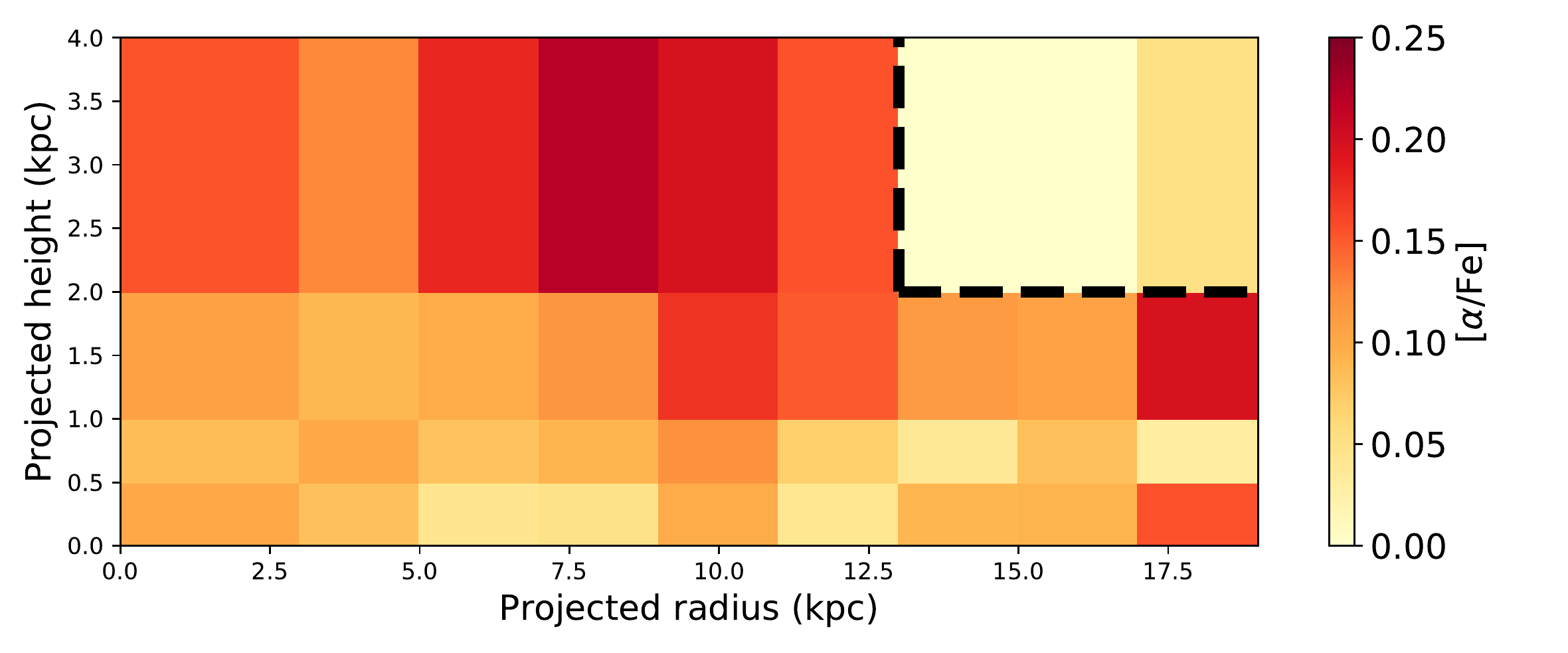}
    \includegraphics[width=3.5in,clip,trim = 10 10 20 10]{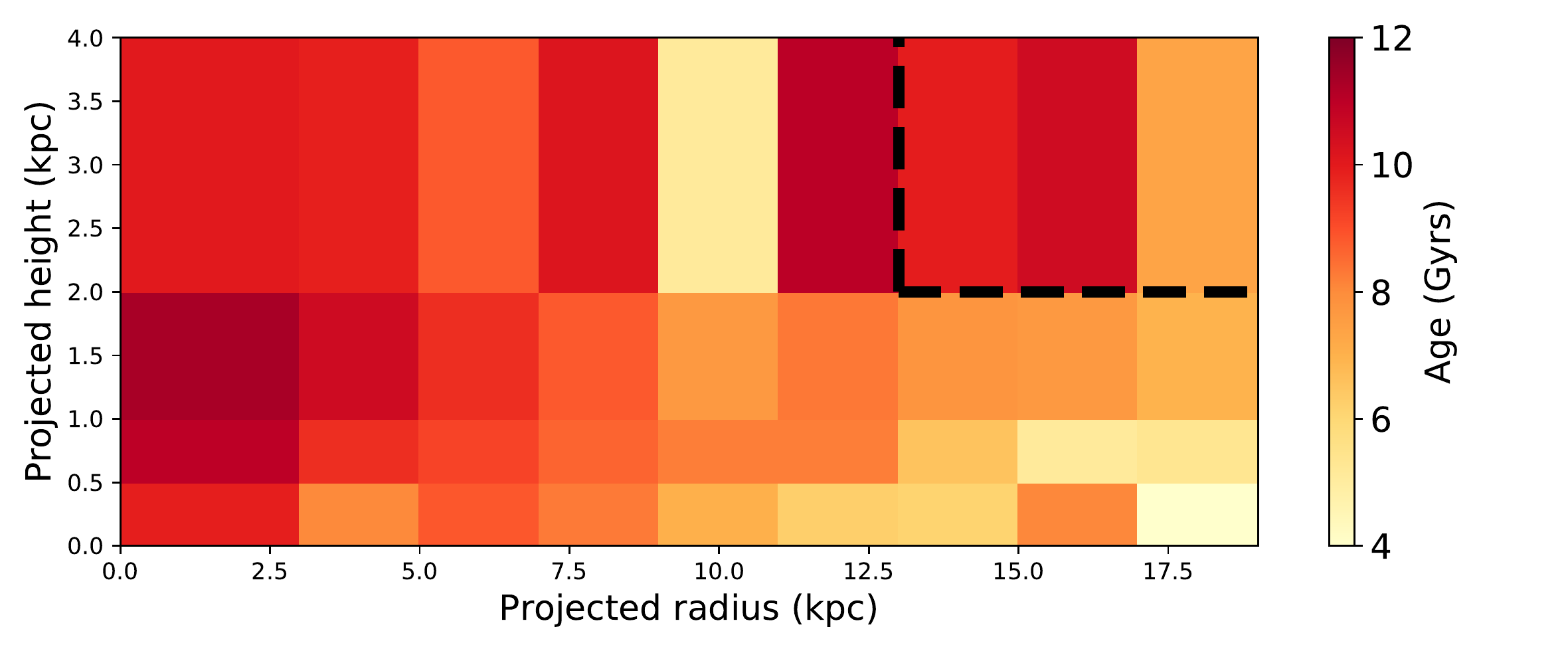}
    \caption{Maps of the mass-weighted average [Z/H], [$\alpha$/Fe] and age (in Gyrs) in UGC 10738. Values have been averaged over all four quadrants of the galaxy. The dashed black line in the upper left of each panel indicates the low S/N region.}
    \label{fig:population_maps}
\end{figure}

To determine the combination of SSP template spectra that best reproduces each observed spectrum we again use pPXF \citep{Cappellari:2004,Cappellari:2012}. In addition to the SSP template spectra described above, we include six emission line templates, including three doublets for nine lines in total (H$\beta$, H$\alpha$, [SII]$\lambda$6717,6731, [OI]$\lambda$6300, [OIII]$\lambda$4959,5007 and [NII]$\lambda$6548,6583) to correct for gas emission and a 10$^\mathrm{th}$ order multiplicative polynomial to account for dust extinction and residual flux calibration errors. pPXF determines the weights of each SSP template required to reproduce the observed spectrum, which represent the fraction of stars (by star-forming mass) with a given age, [Z/H] and [$\alpha$/Fe] in the best fit.

\begin{figure}
    \centering
    \includegraphics[width=3.5in,clip,trim = 10 10 20 10]{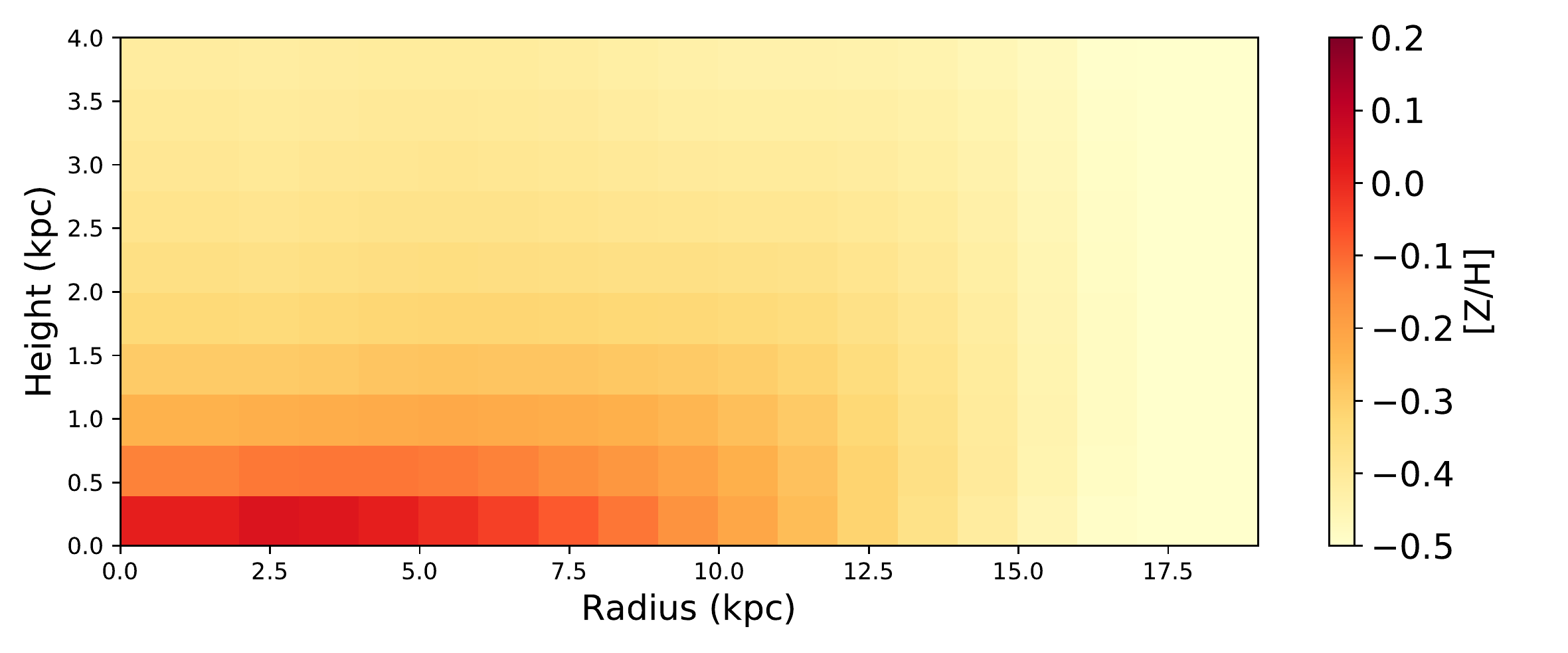}
    \includegraphics[width=3.5in,clip,trim = 10 10 20 10]{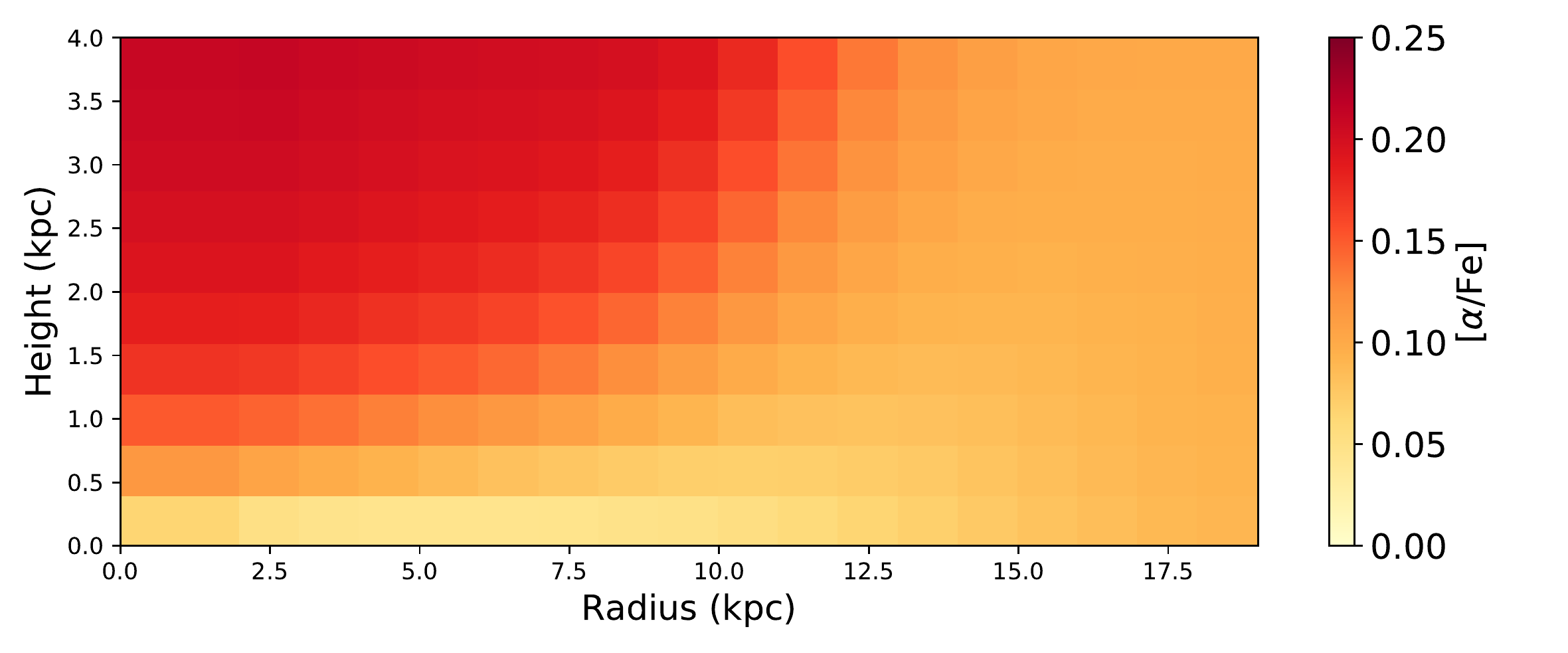}
    \includegraphics[width=3.5in,clip,trim = 10 10 20 10]{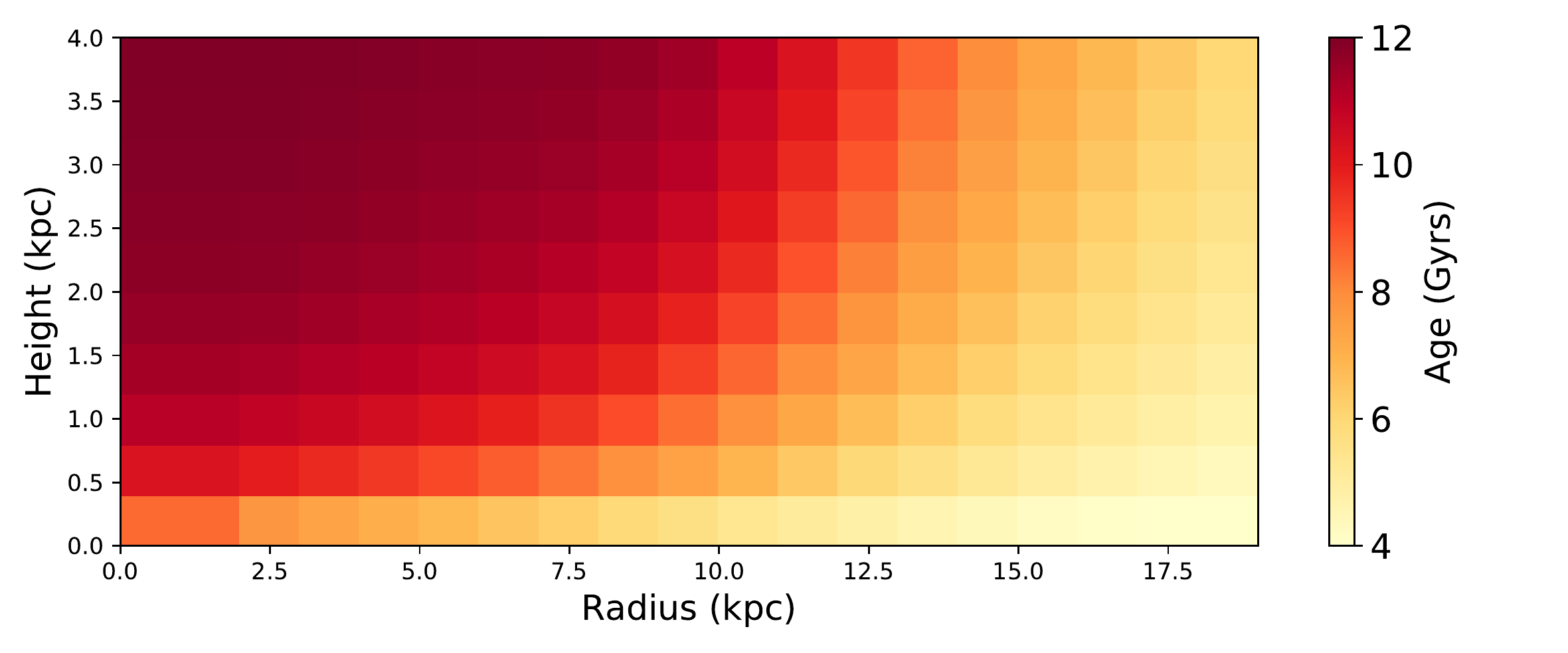}
    \caption{Maps of the mass-weighted average [Z/H], [$\alpha$/Fe] and age (in Gyrs) for the Milky Way, derived from the model presented in \citet{Sharma:2020}, showing the same spatial extent as in Fig \ref{fig:population_maps}.}
    \label{fig:mw_comparison}
\end{figure}

By default pPXF treats all the template spectra independently, which can lead to unphysically spiky star formation and chemical enrichment histories. To mitigate this we apply a second-order regularization to the template weights in the age and [Z/H] dimensions, as described in \citet{Cappellari:2017} (no regularization is imposed in the [$\alpha$/Fe] dimension). We perform an initial fit without regularization to set a baseline goodness-of-fit parameterised by $\chi^2_0$, then gradually increase the regularization until the fit quality is reduced to the point that $\chi^2_\mathrm{final} = \chi^2_0 + \sqrt{2\times dof}$, where $dof$ is number of spectral pixels in the fit. This results in the smoothest star formation and chemical enrichment history that is consistent with the observed spectrum.

An example full spectral fit for the region $5 < |R| < 7$\ kpc and $0.5 < |z| < 1$\ kpc is shown in Figure \ref{fig:pops_example}, with the input (black) and best-fitting (red) spectra shown in the left panel. The mass-weighted fraction of stars with a given age, [Z/H] and [$\alpha$/Fe] is shown in the right panels. For each region we measure the full star formation and chemical enrichment history (i.e. the template weights), and also derive mass-weighted average age and [Z/H] and the fraction of high and low [$\alpha$/Fe] stars.

UGC 10738 has a prominent dust lane through the centre of the galaxy. To determine the impact of dust on our measured stellar population parameters we perform separate population fits to binned spectra from each quadrant at a given radial and height range. We find no significant difference between the [$\alpha$/Fe] values measured above and below the midplane. We find a small, 1 Gyr offset between the stellar ages measured above and below the midplane at the lowest scale heights, $|z| < 0.5$\ kpc, which is not statistically significant given the scatter in ages measured between the different quadrants. At larger scale heights we measure no difference in stellar age. For [Z/H], we find a statistically significant offset for heights $|z| < 1$\ kpc, in the sense that [Z/H] is 0.2 (0.15) dex lower in dust dominated quadrants at heights $|z| < 0.5$\ kpc ($0.5 < |z| < 1$\ kpc). In the dustier regions we probe shallower lines-of-sight into the disk, so we are measuring properties of stars at a larger average intrinsic radius and scale height where we expect [Z/H] to be lower. In summary, we find that dust has a modest effect on the measured stellar population properties at $|z| < 1$\ kpc, but negligible effect at larger heights.

\section{Results} \label{sec:results}

\begin{figure*}
    \centering
    \includegraphics[width=6.5in,clip,trim = 10 10 20 30]{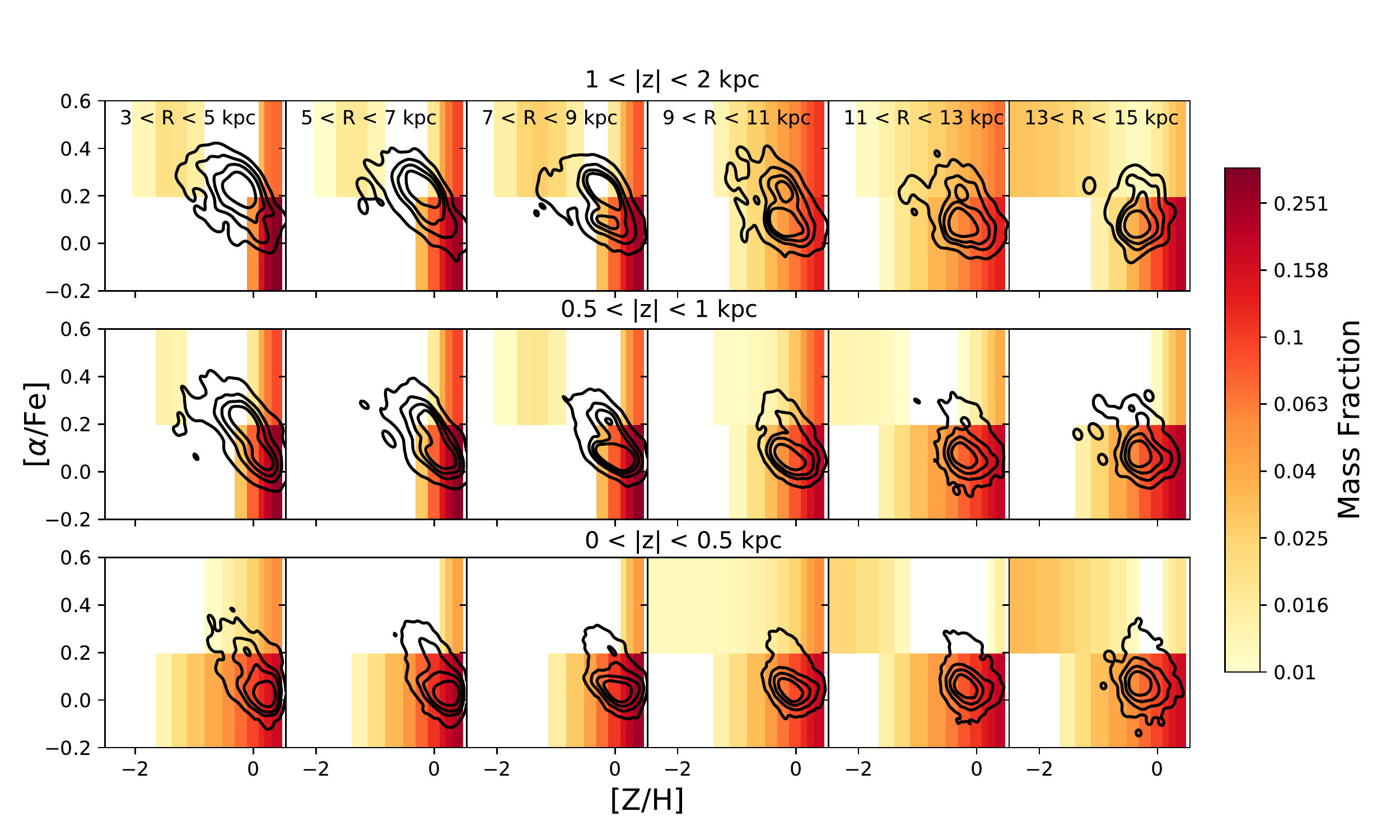}
    \caption{Comparison of the spatially resolved chemistry of UGC 10738 and the Milky Way. Mass fraction of stars as a function of [Z/H] and [$\alpha$/Fe] (summed over age) in bins of projected radius and scale height. Regions are chosen to match those of \citet{Hayden:2015}, though we note the scale radii and height of the two galaxies are likely different. Contours enclose 50, 75, 90 and 99 per cent of Milky Way stars from \citet{Hayden:2015}. We find a high degree of qualitative agreement between the two galaxies, with a low metallicity, alpha-rich component at large scale heights evident in both.}
    \label{fig:alpha_z_regions}
\end{figure*}

In Figure \ref{fig:population_maps} we show the average, mass-weighted age, [Z/H] and [$\alpha$/Fe] of UGC 10738 in the extracted regions. From low to high $|z|$, [Z/H] generally decreases and [$\alpha$/Fe] generally increases at all radii, with the exception of the $|z| < 0.5$\ kpc region which is somewhat lower in [Z/H] than the adjacent height bins, which may be due to dust extinction as described above. 

The mean age of stars increases with height above the plane but declines with radius, such that the oldest stars typically lie in the centre of the galaxy at large heights. At large projected radius and height we find a population with old average ages and low [$\alpha$/Fe], perhaps indicative of a flaring alpha-poor disk, though we note that uncertainties in this region are increased to $\sim 0.15, 0.2, 0.1$ in log(age), [Z/H], [$\alpha$/Fe] respectively \citep[consistent with expectations from][]{Liu:2020}.

Figure \ref{fig:mw_comparison} shows the same quantities as figure \ref{fig:population_maps} but for the Milky Way. The Milky Way values were derived from the chemodynamical model of \citet{Sharma:2020} using a star-forming mass weighting equivalent to that used for the SSP measurements presented here. Unlike figure \ref{fig:population_maps}, these maps show intrinsic spatial regions, so, while the extent of the maps for the two galaxies are comparable they are not exactly equivalent. Despite this we see good qualitative agreement between the spatial variation of the average age, metallicity and [$\alpha$/Fe]. The comparison further improves when noting the relatively larger scale height and radius for UGC 10738 than for the Milky Way. We also find evidence of disk flaring of the young, low-alpha stars in UGC 10738 that is qualitatively similar to that seen in the Milky Way.

In Figure \ref{fig:alpha_z_regions} we show the mass fraction of stars with a given [Z/H] and [$\alpha/$Fe] in the same regions described above. Here we restrict the regions shown to $3 < R < 15$\ kpc and $|z| < 2$\ kpc, the same spatial extent examined in the Milky Way by \citet{Hayden:2015}, again noting that Milky Way regions are in intrinsic radius and height,  whereas for UGC 10738 we have only line-of-sight integrated quantities. The contours indicate the number density of stars in the Milky Way from \citet{Hayden:2015}, transformed from [Fe/H] to [Z/H] following \citet{Salaris:2005}. We see good qualitative agreement between the chemical enrichment pattern seen in the Milky Way and that shown here for UGC 10738. At radii less than $\sim 11$\ kpc and at $|z| < 1$\ kpc, we find the chemical distribution of both galaxies' stars are dominated by a metal rich, alpha-poor component --- the low-alpha thin disk. At the same radii, but $|z| > 1$\ kpc a significant metal poor, high alpha component is present --- the alpha-rich thick disk.

At larger radii qualitative differences between the two galaxies arise, particularly the presence of high-alpha stars in UGC 10738 that are not seen in the Milky Way. This may indicate that UGC 10738's alpha-rich disk is more extended than that of the Milky Way. We also note that in all regions UGC 10738 exhibits contributions from a modest population of high-alpha, metal rich stars that are not seen in the Milky Way. The most likely explanation for the existence of these stars is the lack of intermediate [$\alpha$/Fe] templates available for the full spectral synthesis fitting, which forces pPXF to compensate for e.g. $[\alpha/\mathrm{Fe}] \sim 0.1$\ stars found in the Milky Way by adding in a small mass fraction of $[\alpha/\mathrm{Fe}] = 0.4$ stars in UGC 10738. This ambiguity is likely to be resolved as population templates with intermediate [$\alpha$/Fe] become available.

\section{Discussion} \label{sec:discussion}

\subsection{Extragalactic thick disks}

It is clear that structural thin and thick disks are not uncommon in the Universe \citep{Yoachim:2006,Martinez-Lombilla:2019}, potentially being found in the majority of disk galaxies \citep{Comeron:2018}. This study demonstrates that chemically thick disks, i.e. galaxies that contain high- and low-[$\alpha$/Fe] populations that have different spatial distributions, are also found in other Milky Way-like galaxies. From this object, and a small number of lenticular galaxies in the Fornax cluster \citep{Pinna:2019,Pinna:2019b} which show similar enrichment patterns, we cannot yet say whether chemically distinct thick disks are rare, common or ubiquitous, however we do now know that this structure is not unique to the Milky Way.

The spatial distribution of [$\alpha$/Fe] rich and poor stars in UGC 10738 is qualitatively similar to that seen in the Milky Way, as shown in Figure \ref{fig:alpha_z_regions}. At present, empirically-based SSP models have limited resolution in the [$\alpha$/Fe] axis, making a more quantitative comparison challenging, however the `knee' in the [$\alpha$/Fe] -- [Fe/H] distribution in UGC 10738 may occur at somewhat lower values of [Fe/H] than in the Milky Way. The metallicity of the knee is indicative of the efficiency or rate of star formation during the formation of the thick disk \citep{Matteucci:1990,Pagel:1997,Andrews:2017} --- lower star formation efficiency, or lower star formation rates result in less metal-enriched thick disk stars, indicating that thick disk formation was less efficient in UGC 10738 than in the Milky Way.

\subsection{Implications for the formation of the Milky Way}

There are two leading explanations for the apparent chemical bimodality observed in the Milky Way; a similarly bimodal formation history \citep{Chiappini:1997,Haywood:2016}, or as a natural consequence of how the $\alpha$ elements are produced \citep{Schoenrich:2009,Sharma:2020,Vincenzo:2020}. The bimodal formation history scenario consists of an early starburst-like phase followed by rapid quenching and a more gradual extended period of star formation. This initial starburst may be triggered by an interaction \citep{Grand:2020}, or may occur naturally as a result of the gas accretion history from the cosmic web. If this assumed star formation and merging history is a necessary prerequisite to form a bimodal [$\alpha$/Fe] distribution, this would imply that all galaxies with bimodal [$\alpha$/Fe] distributions should have experienced similar histories, including a merger-triggered star burst phase $\gtrsim 7$\ Gyrs ago. 

The alternative scenario does not require a break or pause in star formation, instead consisting of a smooth star formation history with the bimodality arising from early, very high star formation rates producing a significant number of high [$\alpha$/Fe], low [Z/H] stars before rapid enrichment of gas from Type Ia supernovae causes a sharp transition to the low [$\alpha$/Fe] sequence \citep{Schoenrich:2009}.

While the chemical enrichment patterns of the two scenarios are, by construction, extremely similar, their relative likelihoods in a large population of Milky Way-like galaxies should be very different. Galaxies that experience a sharp pause in star formation at the required redshift, especially one triggered by a substantial merger event, are likely to be relatively rare; \citet{Evans:2020} found only 5 per cent of Milky Way-mass galaxies have similar assembly histories to the Milky Way (undergoing an early, massive accretion event $\sim 10$ Gyrs ago) in the large-volume cosmological hydrodynamical EAGLE simulation, with \citet{Mackereth:2018} finding chemical bimodalities to be rare in the same simulation \citep[however][find gas-rich mergers to be present in all of a small sample of high-resolution Milky Way analogue simulations]{Buck:2020}. In contrast, Hayden et al. (in prep.) find chemical bimodalities to be ubiquitous in Milky Way-like galaxies in a chemodynamical simulation with chemical enrichment chosen to reproduce the `smooth' scenario described above. These predictions strongly suggest that the relative prevalence of chemical bimodalities in MWAs has the power to distinguish between the likely origin of the Milky Way's own chemical bimodality. 

This tension is further enhanced once early-type disk galaxies with [$\alpha$/Fe]-enhanced thick disks are included in the population of present day galaxies with similar formation histories \citep{Pinna:2019b,Poci:2019,Poci:2021}. That early-type disk galaxies are found to contain similar abundance patterns as found in the Milky Way (i.e. increased mean [$\alpha$/Fe] off the plane of the disk) is unsurprising. Such galaxies represent one plausible end point for the Milky Way's evolutionary history, suggesting a shared and generic evolutionary pathway for disk galaxies. The Milky Way is often identified as a so-called `green valley' galaxy \citep{Mutch:2011}, a galaxy already undergoing a transition to the red sequence. When fully quenched (and assuming no dramatic structural changes in the mean time) the Milky Way will likely resemble a lenticular galaxy similar to FCC 170 \citep{Pinna:2019}. 

If bimodal [$\alpha$/Fe] distributions and chemically distinct thick disks are found to be generic features of Milky Way-like galaxies this may pose a challenge for the merger-triggered starburst theory of the origin of the Milky Way's own thick disk. A ubiquitous feature is unlikely to be the result of a highly stochastic process like a major merger, being more likely to arise from secular processes or the integrated history of accretion and interaction with the interstellar medium \citep{Sharma:2020,Vincenzo:2020}. The lack of chemical bimodalities in large-scale cosmological simulations such as EAGLE \citep{Mackereth:2018} may indicate those simulations do not sufficiently resolve the ISM to reproduce disk structures accurately \citep{Lagos:2018}, or include sufficiently sophisticated enrichment physics \citep{Kobayashi:2020}.

Characterising the relative frequency of chemically-distinct disk components in Milky Way-like galaxies provides an important test of these two competing theories of thick disk formation. This study demonstrates that, with sufficiently high quality IFS data it is possible to meaningfully compare the spatially-resolved chemical distributions of external galaxies to that of the Milky Way. The next step is to expand the sample of Milky Way galaxies with the necessary observations, providing the first measurement of the frequency of chemically distinct components in a statistically significant sample of Milky Way-like galaxies. 

\acknowledgments
We thank the anonymous referee for their feedback on improving this article. We would like to thank Eric Emsellem for assistance with the MUSE data reduction. NS and JvdS acknowledge support of Australian Research Council Discovery Early Career Research Awards (project numbers DE190100375 and DE200100461, respectively) funded by the Australian Government. RMcD is the recipient of an Australian Research Council Future Fellowship (project number FT150100333).

%

\vspace{5mm}
\facilities{ESO(MUSE), Pan-STARRS}


\software{astropy \citep{2013A&A...558A..33A};
            MUSE Data Reduction Software v2.8.1 \citep{2016ascl.soft10004W};
            ZAP \citep{2016MNRAS.458.3210S};
            pPXF \citep{Cappellari:2012}.
          }




\bibliography{muse_alpha_disks}{}
\bibliographystyle{aasjournal}



\end{document}